# A Review on Recent Energy Harvesting Methods for Increasing Battery Efficiency in WBANs

Hossein Yektamoghadam, Amirhossein Nikoofard, *Member, IEEE*, Fatemeh Pourhanifeh Doust, and Mehdi Delrobaei, *Senior Member, IEEE*

*Abstract*— Today, technology development has led humans to employ wearable and implantable devices for biomedical applications. An important research issue in this field is the wireless body area networks (WBANs), which focus on such devices. In WBAN, using batteries as the only energy supply is a significant challenge, especially in medical applications. Charging the batteries is a problem for patients who use WBAN. Replacing the battery is not very difficult for wearable devices, but implantable devices have different conditions. The use of batteries in implantable devices has many problems, including pain and costs due to surgery, mental stress, and lack of comfort. Batteries' life depends on their type, operation, the patient's medical condition, and other factors. This paper reviews recent energy harvesting methods for battery recharge in WBAN's sensors. Moreover, we provide future research directions on energy harvesting methods in WBANs. Therefore, active research fields such as reinforcement learning (RL) and distributed optimization in WBAN applications were investigated. We strongly believe that these insights will aid in studying and developing a new generation of rechargeable sensors in WBANs for fellow researchers.

## I. INTRODUCTION

Wireless sensor networks (WSN) are a group of spatially dispersed and dedicated sensors responsible for monitoring and recording data from the environment in a central location. WBANs are a subset of the WSN focusing on home and hospital monitoring of patients or for monitoring the elderly in nursing homes, avoiding unnecessary hospitalization of the patient, and similar cases have been used to increase public health. An example of non-medical applications of WBANs is their use in professional sports to improve athletes' fitness. In current professional sports, it is essential to have control and care technologies for a professional athlete to record and analyze an athlete's physical condition. WBANs can help athletes and coaches improve their mental and physical fitness. WBANs have a more vital role in health applications. The need for intelligent monitoring of patients' conditions arises to control them continuously by monitoring vital signs. WBANs can have effective control over human health status. Consequently, emergency treatment is applied in emergencies like abnormal heart rate changes [1]–[7].

Hossein Yektamoghadam is with the School of Electrical and Computer Engineering, College of Engineering, University of Tehran, Tehran, Iran (e-mail: Yektamoghadam@ut.ac.ir).
Amirhossein Nikoofard is with the Department of Electrical Engineering, K. N. Toosi University of Technology, Tehran, Iran (e-mail: a.nikoofard@kntu.ac.ir; Delrobaei@kntu.ac.ir).
Fatemeh Pourhanifeh Doust is with the Department of Electrical Engineering, Faculty of Engineering, University of Guilan, Rasht, Iran (e-mail: Fatima.pourhanife@gmail.com).
Mehdi Delrobaei is with the Department of Mechatronics, Faculty of Electrical Engineering, K. N. Toosi University of Technology, Tehran, Iran, and the Department of Electrical and Computer Engineering, Western University, London, ON, Canada (e-mail: delrobaei@kntu.ac.ir).

The use of batteries in implantable devices has many problems, including pain and costs due to surgery, mental stress, and lack of comfort. Batteries' life depends on their type, operation, the patient's medical condition, and other factors. The sensors in WBANs are constantly collecting and transmitting data from the environment or the body. Therefore, these sensors must be continuously rechargeable, and the idea of energy harvesting has been proposed to solve this issue. Replacing optimized methods to supply WBAN sensors instead of batteries can promote controlling and improving patient conditions [8]–[10]. Energy harvesting is defined as the collection of small energies that may be wasted in any way, such as light, sound, heat, vibration, or movement. Energy harvesting from WSNs has been harvested through environmental energies such as the sun, wind, sea, and waves. Ecological energy is a desirable energy source, enabling sensor nodes to be self-sustaining, potentially leading to an extremely long lifespan. Energy harvesting from WBANs considers ambient and human body energy as two sources [11]–[13].

Energy harvesting from ambient sources is the most common method used in different applications of energy harvesting in WBANs. Ambient sources can harvest considerable energy, which can power a range of wearable and implantable applications if they are effectively harnessed. Light is one of the most mature sources, which can be used outdoors and indoors [14]–[16]. On the other hand, energy harvesting from the human body is another way to generate energy in WBAN. Energy harvesting from ambient methods used for WSN and WBAN. Still, human body methods are often proposed for WBAN applications' energy supply. Therefore, more work has been done on ambient techniques and technologies than human body methods. Energy harvesting from the human body takes the appropriate energy from different sources of the human body, such as glucose or muscle movement. It can then supply energy demand for healthcare applications of wireless systems [17]–[19].

WBANs are comprised of a control unit and various sensors, which can either be implanted in the body or worn externally as body-worn or body-patched devices. These sensors can communicate wireless with each other or with a control unit. In the WBAN implant, a medical device that is implanted in the body sends its measured biological data to a designated hub using at least one on-body device to meet specific size, service quality, and power consumption requirements. Because these sensors consume low energy, this amount of energy can be provided with the help of ambient and the human body. Therefore, researchers are trying to provide more comfort and reliability for patients who use WBAN devices [20]–[22].

In order to use the harvested energy, it is necessary to satisfy the hard and soft constraints. In the first step, the energies obtained from the ambient and the human body should be harvested with the help of existing technologies. The harvested energy should be stored in the battery using power electronic devices in the second step. Also, energy saving is a method that can reduce energy consumption and increase efficiency. Energy saving is a set of approaches and algorithms that can increase battery life. A graphical overview of this paper is shown in Fig. 1. Also, Fig. 1 describes the energy cycle architecture for the WBAN application.

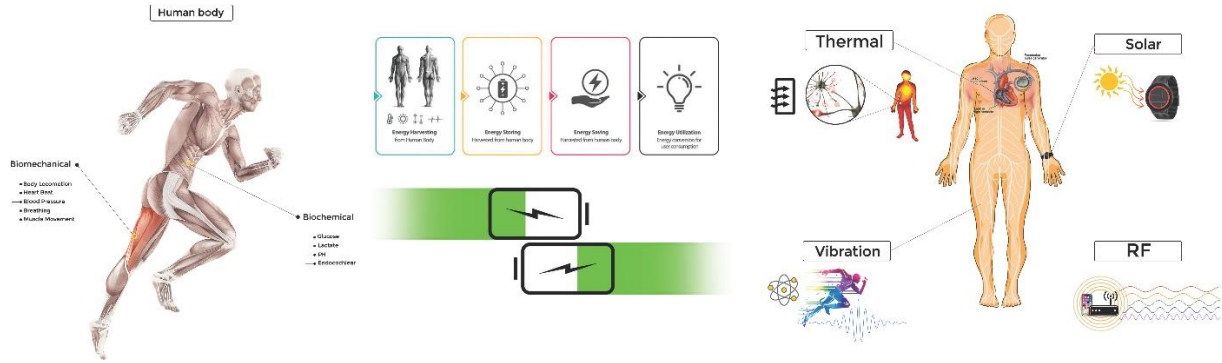

Fig. 1. Overview of energy harvesting methods for WBAN application.

This survey paper investigates the energy harvesting of WBANs and their future challenges. A comprehensive overview has been presented with a deep focus on a recent technical overview of the state-of-the-art technologies of energy harvesting in WBANs. This paper has different survey sources of energy harvesting from the human body, their characteristics, methods, classification, and types. Moreover, future research directions for energy harvesting in WBANs have been presented, and critical challenges have also been identified.

## II. ENERGY HARVESTING FROM AMBIENT

Energy can be harvested from different sources. In WBANs, it is impossible to charge the battery through the ambient for implanted sensors in the body. However, this method can be suitable for wearable sensors. A detailed architecture of energy harvesting methods for increasing battery efficiency in WBANs is present in Fig. 2. The sources are divided into two groups the ambient and the human body. There are many similarities between WSNs and WBANs in ambient methods. Therefore, the existing technologies are close to identical. In the following, we will explain the different types of energy harvesting from ambient methods.

### A. Solar energy harvesting

Solar energy harvesting in WBANs involves utilizing solar cells to convert sunlight into electrical energy to power wearable or implantable devices within the body area network. Solar energy is an unlimited source of energy, and a popular option due to its non-polluting for the environment. Photovoltaics are the most mature method to harvest energy from solar. This technology can convert solar energy into electricity through the use of photovoltaic properties semiconductors. The primary material of photovoltaic cells is silicon. Therefore, there is a potential difference between the negative and positive electrodes when the sun shines on them to create an electric current. Photovoltaic technology has a more tremendous advantage than other methods due to its high efficiency. In typical cell types, the efficiency is about 30%, with a power density of 100 to 1000 micrograms per square centimeter indoors and 100 megawatts per square centimeter outdoors for single crystal cells. However, the efficiency of these cells is less than the available potential because these cells cannot absorb photons with long and short wavelengths in the full spectrum of light [23], [24].

Researchers are trying to use nanotechnology to topography's change of these cells, allowing them to absorb ultraviolet wavelengths. Photovoltaic technology has been successful commercially in supply energy on a large and small scale. Solar energy harvesting from WBANs is a recharge method of the battery to sensors energy supply. Small size solar panels are conveniently connected to low-energy and rechargeable battery circuits. Therefore, they create WBAN nodes with an infinite network life. This technology can be divided into three parts: solar, as an energy source, energy harvesting unit including photovoltaic cells, and WBANs power supply battery units. Energy harvesting technology is often shared between the WSNs and WBANs. Energy harvesting systems should be designed to convert and store solar energy efficiently. This involves selecting appropriate solar cell technologies, power management circuits, and energy storage solutions like batteries or supercapacitors [25]–[27].

### B. Radiofrequency energy harvesting

Radiofrequency (RF) energy harvesting in WBANs involves capturing and converting ambient RF signals into electrical energy. This technology provides an alternative or complementary approach to traditional battery-based power sources, enabling energy autonomy for WBAN devices. RF energy harvesting is a promising solution to achieving an acceptable power level for RF waves in places without sunlight. However, energy harvesting from this method has many challenges. RF waves are transmitted through wireless channels, and they are affected by various phenomena such as shadows. Therefore, the power amount received by an antenna from an RF energy harvester is randomized, and the systems designer must consider these constraints [28]–[30].

Ambient RF signals are generated by various wireless communication systems, including Wi-Fi, cellular networks, and other radiofrequency sources. The energy harvested from RF signals is typically low, so efficient energy conversion and storage mechanisms are crucial. The design of antennas plays a critical role in RF energy harvesting efficiency. Antennas must be tuned to specific frequencies of interest to capture and harvest RF signals effectively. Compact and flexible antenna designs are desirable for integration into wearable devices. RF energy harvesting circuits often include rectifiers, voltage multipliers, and power management components. Harvested RF energy is typically intermittent, and energy storage elements such as batteries or supercapacitors are used to store and provide a stable power supply to WBAN devices during periods of low or no RF signal availability [31]–[33].

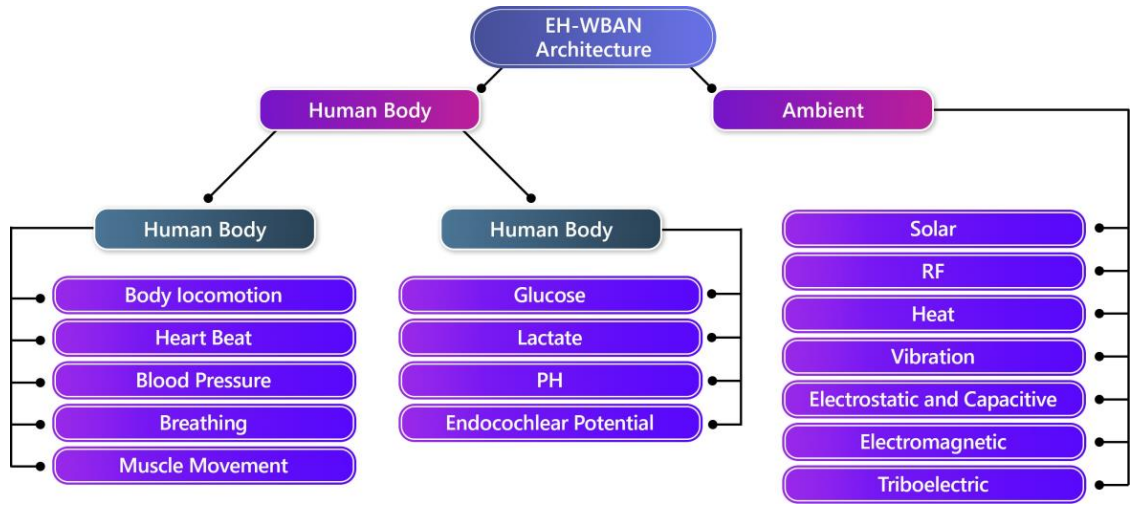

Fig. 2. Architecture of energy harvesting methods for increasing battery efficiency in WBANs.

### C. Thermal energy harvesting

Thermal energy harvesting from the human body or other sources near the body represents a promising alternative method to energy harvesting due to external factors in WBAN. Thermal energy harvesting is based on the temperature differences between two levels to generate electrical energy. Therefore, the device's performance highly depends on human activity and ambient factors. Generally, the Thermal energy harvesting follows Seebeck, Peltier, and Thomson effect [34], [35].

In 1821, Tomas Seebeck discovered the Seebeck effect, which involves heating one surface of the Peltier element, cooling the other surface, and producing an electrical current at its end terminals. The Seebeck and Peltier effects are the same physical process with different manifestations. The Peltier effect is heating or cooling at an electrified junction of two dissimilar conductors. Jean Charles Athanase Peltier, the French physicist, discovered this effect in 1834. The difference between materials in materials by this effect and other materials caused the Seebeck factor to be constant in other materials. Therefore, a spatial gradient in temperature can result in a rise in the Seebeck coefficient. A continuous Peltier effect is created if a currency has driven through this gradient. This Thomson effect was predicted and later observed in 1851 by William Thomson. The advantages of Thermal energy harvesting con- verters include solid-state and not having an active part, low noise pollution, high reliability, small weight and size, and comfortable orientation. Conversely, low efficiency has been a significant disadvantage in these devices that engineers are trying to increase [36]–[38]. A schematic of the thermoelectric generator device is shown in Fig. 3.

### D. Vibration energy harvesting

A vibration Energy Harvesting System can harvest energy from human motion and environmental vibrations. Human motion analyzes the vibrations generated by human motion, focusing on gait, limb movement, and daily activities. On the other hand, environmental vibrations explore the potential of harvesting energy from external environmental sources, such as vibrations from machinery or infrastructure. Vibration energy harvesting systems include transducer design, power conditioning circuitry, and Integration with WBAN devices [39], [40]. The harvesting of energy from vibration is one of the famous methods of energy harvesting, which is mainly applied with the piezoelectric approach. The piezoelectric energy harvesting method is based on the properties of materials that can generate charge and electric field when the mechanical force is applied, called the piezoelectric effect. Piezoelectric converters can have different shapes; they are suitable for many applications and similarly change a crystalline substance's conditions. The Piezoelectric energy harvesting method is commonly used to harvest energy from other sources in the human body. Walking and running are essential factors in generating energy from this ambient source. Energy harvesting from moving or running can generate significant electrical energy and is a current research interest subject for many researchers [41], [42].

### E. Electrostatic and Capacitive energy harvesting

Electrostatic energy harvesting is a method to gather the amount of energy from kinetic energy sources. Accordingly, electrostatic energy harvesting devices use moving plates of variable capacitors charged and discharged periodically. The structure of this system works through oscillation in the electric field and converts the mechanical energy into electrical energy [43]. On the other hand, in the electrostatic and capacitive energy harvesting method, when one of the concealer masses intensifies oscillates, the vibration amplitude is large enough to connect and collide with another mass and provides strong mechanics between two subsystems. Therefore, this operation can create a suitable energy source from vibration sources in a wide frequency range. Microelectromechanical systems include wearable wireless sensors, and pacemakers are the most common applications of this technology [44].

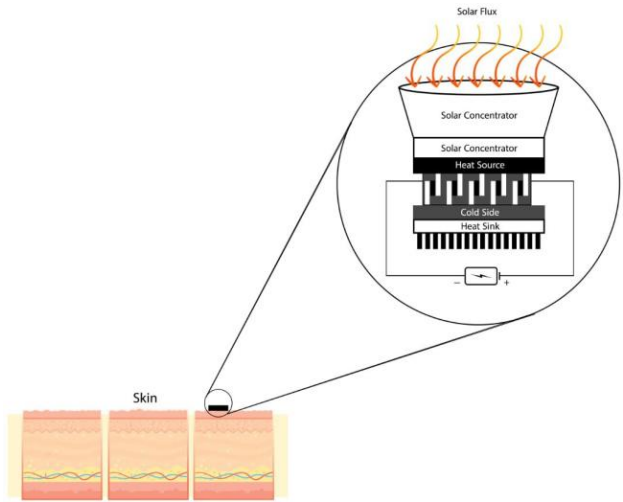

Fig. 3. A schematic of the thermoelectric generator device. This device includes a solar concentrator, heat source, cold side, and heat sink.

They are resistant to radiation, susceptible to vibrations, and have good stability at various temperatures. Moreover, they include military and space applications. Capacitive energy harvesting involves placing a conductive shell between the ground and a conductor to form a capacitive voltage divider. Independent implementation of load and not being affected by weather conditions are the most important advantages of this method. Researchers have shown this method can produce an output power of 0.5 watts when the voltage between the conductive shell and the collector is 5 kV.

### F. Electromagnetic induction energy harvesting

Electromagnetic induction is a method that includes a conductor, usually a coil that breaks flux lines, a magnet that creates an EMF, and a current on the power supply coil. Electric rotation machines, wind, and waves are examples of this method. The simple operation of magnetic flux lines broken by a conductor can be used in a wide range of energy-harvesting applications. In many cases, the electromagnetic induction energy harvesting method and other methods, such as the piezoelectric method, are used for more energy harvesting efficiency [45], [46].

Ambient magnetic field Sources include environmental sources and human-generated Fields. Environmental sources identify and characterize environmental magnetic field sources, including natural and artificial contributors. Moreover, human-generated fields investigate magnetic fields produced by physiological activities, focusing on their suitability for energy harvesting. Coil Size Placement and Variability in Magnetic Fields are the main Challenges in electromagnetic induction energy harvesting in WBANs [47].

### G. Triboelectric energy harvesting

A triboelectric nanogenerator is an energy-harvesting device that can generate electricity from external mechanical energy through triboelectric and electrostatic induction. However, these nanogenerators also have broad applications for energy harvesting from the human body, which we will discuss in the next section. Firstly, this new type of nanogenerator was demonstrated in Prof. Zhong Lin Wang's group at the Georgia Institute of Technology in 2012. This generator transfers electrons between two interfering surfaces by non-identical materials [48]. Triboelectricity has been viewed as detrimental due to its tendency to cause electronic circuit malfunctions, fires, and discomfort when not properly managed. These risks arise from the excessive voltages of triboelectricity, which can reach tens of thousands of volts, especially in dry conditions. Still, these hazards are less dangerous at low voltages used in WBAN [49]–[52]. A summary of the latest representative research in ambient energy harvesting in WBANs is shown in Table 1.

## III. ENERGY HARVESTING FROM THE HUMAN BODY

Energy harvesting from the human body in WBANs refers to capturing and converting various forms of energy produced by human activities into electrical power. This harvested energy is then utilized to power electronic devices and sensors embedded in wearable or implantable devices that form a wireless network on or in the human body. Energy harvesting from the human body is another way to generate energy in WBAN. Energy harvesting from ambient methods used for WSN and WBAN. Still, human body methods are often proposed for WBAN applications' energy supply. Therefore, more work has been done on ambient techniques and technologies than human body methods. Energy harvesting from the human body takes the appropriate energy from different sources of the human body, such as glucose or muscle movement. It can then supply energy demand for healthcare applications of wireless systems [53], [54]. The structure of WBAN sensors in the human body is shown in Fig. 4.

### A. Biochemical energy harvesting

Biochemical methods generate electricity from the body's chemicals. Biofluids in the human body contain various active substances and enzymes that supply energy to the body under certain conditions. These chemical compounds can be obtained by harvesting electrochemical reactions. The amount of energy harvested from biochemical methods in the body depends on the person's age. It mainly de- depends on health and the daily consumption of nutritious food. A distinct feature of this method is energy harvesting every time, and the body maintains a good source of potential. On the other hand, implantable devices that harvest energy from biological sources and are based on electrochemical transducers are currently receiving close attention. It can be a promising method to harvest and utilize energy from the body to activate wireless nodes in WBAN [55]–[60]. Some crucial sources in this method are:

• Glucose: Implantable glucose fuel cells are a novel technique to use autonomous energy supply for WBAN nodes, especially in medical applications based on the electrochemical reaction of oxygen and glucose. Glucose energy harvesting in WBANs involves the utilization of glucose, a sugar present in the human body, as a potential source of energy to power electronic devices and sensors within wearable devices, forming a wireless network. This concept taps into the human body's metabolic processes to generate electrical power for various applications. Using a spatial electrochemical reaction, the cell converts endogenous matter and oxygen into electricity in this method. These systems are constantly replenished with new body fluid reactors, which is unlike batteries. Theoretically, the activity of these cells to produce energy is unlimited as long as there is a constant reactive source [61]. In the laboratory and research stage, researchers have created new ideas from this energy harvesting method [62], [63].

TABLE I
SUMMARY OF SOME LATEST REPRESENTATIVE RESEARCH IN AMBIENT ENERGY HARVESTING IN WBANS

| Ref. | Source\Technology | Method | Strategy | Main Work | Output |
|---|---|---|---|---|---|
| [31] | RF Signal | Transmission Power Adjustment Algorithm | Time switching | A MAC protocol for RF Energy Harvesting that operates across multiple layers to improve efficiency | Outstanding performance for extended periodic health monitoring purposes |
| [32] | RF Signal | Relay selection | Cooperation transmission | A method for uploading data in RF energy harvesting powered WBANs using a cooperative-based approach | Outperforms the typical single-hop approach (in a specific example: packet reception rate improvement ranges from 4.9% to 7.8%) |
| [33] | RF Signal | TDMA-based Self-Adaptive MAC schedulling Algorithm | Time allocation | A new MAC protocol that combines kinetic and wireless energy harvesting technologies for EH-WBAN | The efficiency of the program in terms of life cycle and energy usage |
| [36] | Heat | Thermoelectric generator approaches | Comparison | Direct your attention to the important connection between the thermal harvester and the power conditioning circuitry and make a comparison | achieve an average power of 260 $\mu W$ ($\mu TEG$) to 280 $\mu W$ (mTEG) and power densities of 13 $\mu W cm^{-2}$ ($\mu TEG$) to 14 $\mu W cm^{-2}$ (mTEG) for systems worn on the human wrist |
| [39] | Human Activity (Vibration) | Analytical method | Fabrication | The proposal outlines a design for an affordable WBAN that achieves extremely low energy consumption | Obtaining outstanding performance outcomes, especially in terms of wearability, and battery life |
| [40] | Multi-joint movements (Vibration) | Physical dimensions and manufacturing | Fabrication | Develop a wearable triboelectric nanogenerator that utilizes contact-separation direct current to generate power for a self-powered WBAN | Self-sustained wireless monitoring of bodily movements in real-time, powered by low-frequency human daily activities |
| [42] | Human walking (Vibration) | Characterization methods | System-level | The creation and analysis of a portable energy harvesting powered wireless sensing system to tackle the obstacles associated with energy harvesting | The speed of walking experienced a rise from 3 to 7 kilometers per hour, while the power output also saw an increase from 1.9 ± 0.12 to 4.5 ± 0.35 mW |
| [46] | Hybrid vibration and wind (piezoelectric and electromagnetic) | Architecture, fabrication, and characterization of the harvester | Fabrication | A new hybrid bridge energy harvester that utilizes both piezoelectric and electromagnetic conversion methods has been developed | The hybrid bridge energy harvester functions at three distinct low-frequency resonant modes, with frequencies spanning from 11 to 45 Hz, specifically centered around 11, 38, and 43 Hz |
| [47] | Human motion (electromagnetic) | Architecture and fabrication | Fabrication | An energy harvester that combines electromagnetic and triboelectric technologies to achieve high efficiency by converting vibration into rotation | The harvester's active power output remains consistently at around 300 mW while jogging, surpasses 800 mW during sprinting, and is capable of consistently charging a smart band with a rated power of 400 mW |
| [48] | Human body (triboelectrification effect) | Atatistically and systematically | Review | The suggestion is to introduce a concept for an Internet of Things (IoT) system based on triboelectric nanogenerators, and to provide a thorough examination of this IoT system from the perspectives mentioned above | Providing significant perspectives on the advancement of triboelectric nanogenerator-based human body Internet of Things systems |

• Lactate: The complete electrochemical oxidation of lactate is a new method to harvest energy from fuel cells. Lactate energy harvesting in WBANs involves converting chemical energy stored in lactate, a byproduct of cellular metabolism, into electrical energy to power electronic devices and sensors within wearable devices forming a wireless network. This concept leverages the presence of lactate in bodily fluids, such as sweat, as a potential and renewable energy source. Commercially available enzymes can readily oxidize lactate present in abundant levels in human sweat. Sweat lactate is a biomarker known for exercise intensity and muscle activity. During non-invasive body studies conducted by researchers, the density of power obtained from lactate is directly related to sweating in people with different fitness levels [64], [65]. Lactate energy harvesting in WBANs holds promise for applications in sports science, fitness monitoring, and healthcare. It offers a unique approach to simultaneously monitor physiological parameters and generate energy during activities, addressing the need for sustainable power sources in wearable technologies. Ongoing research aims to enhance the efficiency and practicality of lactate energy harvesting systems for real-world applications.

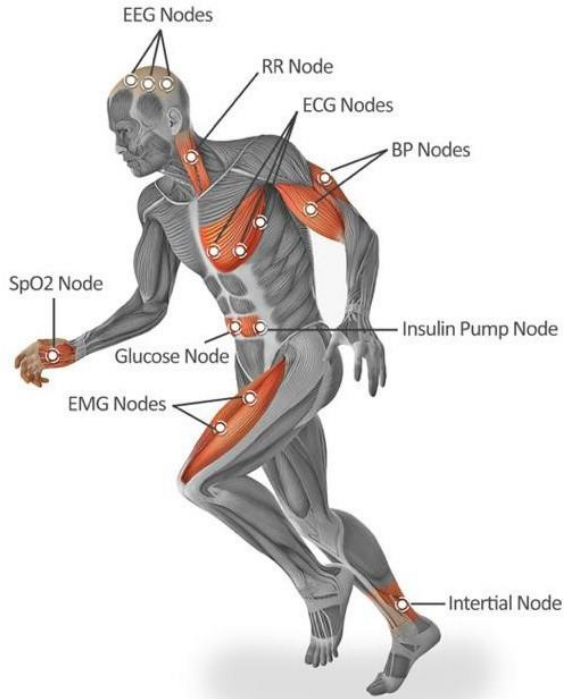

Fig. 4. The structure of WBAN sensors in the human body; The arrangement of the human body sources can supply the sensors. For example, heart beat energy can recharge ECG sensors.

• Potential hydrogen (PH): PH is another energy source suitable for producing very low-level voltages. PH is a measure of the acidity or alkalinity of a solution. PH levels in the human body are regulated within a narrow range, and variations can occur in different bodily fluids. Metal oxide PH electrodes work based on the potential of the electrode measured versus the reference electrode, normally Ag/AgCl, due to the balance between the slightly dissolved salt and the saturated solution. The amount of output voltage depends on the amount of output of thermodynamic solubility. Therefore, solid-state electrodes are easier to miniaturize than glass electrodes and have a faster reaction time [66], [67].

• Endocochlear Potential: The endocochlear potential refers to the electrical potential in the endolymph of the cochlea, the part of the inner ear responsible for hearing. It plays a crucial role in the process of transducing sound vibrations into electrical signals that the brain can interpret. Another viable energy source is the endocochlear potential (EP) by output about 70–100 mV [68]. This method usually involves an electrochemical battery that is actively maintained and typically used to harvest energy near the ears, brain, and eyes; therefore, there is a high sensitivity to exploitation. This method is one of the novel and promising methods expected to receive more attention in the future [69], [70].

### B. Biomechanical energy harvesting

Biomechanical energy harvesting from the human body presents alternative renewable energy supplied for batteries in wireless devices. The key feature of this device is that the power generation adds only a minimal extra effort to the user. In general, energy harvesting in this method is classified into two voluntary and involuntary categories. Voluntary actions involve spontaneous activities related to body movement, walking, exercising, *etc*. Automatic refers to the body's continuous motions, such as breathing, heartbeat, blood pressure, and other muscle movements, regardless of a person's intent [71]–[78]. Some examples of biochemical energy harvesting methods are described in more detail below:

• Body locomotion: Body locomotion energy harvesting means recovering energy from body movement, which worldwide have studied for the past two decades. Body locomotion energy harvesting in WBANs involves capturing and converting the energy generated by body movements into electrical energy that can be used to power devices or sensors within the WBAN. This concept leverages the human body's movements, such as walking, running, or other physical activities, to generate power for electronic devices attached or implanted in the body. Efficient energy harvesting from body movement requires a small, lightweight device that effectively converts intermittent into electricity. To achieve this goal, Linear, Piezoelectric, motion-tracking systems, pressure sensors, and rotary electromagnetic generation methods are often utilized to analyze human locomotion energy harvesting. Body locomotion energy harvesting provides a continuous and sustainable power source for wearable and implantable devices, reducing the need for frequent battery replacements or external power sources. This approach aligns intending to make WBANs more self-sufficient and less dependent on traditional power supplies, particularly in applications related to healthcare monitoring and personalized medicine [79]–[82].

• Heartbeat: Energy harvesting from heartbeat vibrations can be used to power a leadless pacemaker. This approach harnesses the pulsatile motion of the chest or other relevant body parts to generate power for electronic devices within the WBAN. The process typically involves using energy harvesting devices, such as piezoelectric materials or electromagnetic generators, strategically placed on or near the chest to capture the mechanical vibrations associated with each heartbeat. These mechanical vibrations are then converted into electrical energy through transduction mechanisms, such as the piezoelectric effect or electromagnetic induction. More than 1.8 billion heart cycles over a 70-year life with an average heart rate of 70 beats per minute show that the heart muscle is enduring. This makes it a reliable power source that can supply energy to medical implants such as pacemakers. According to the latest research, the consumption of pacemakers has been reduced to 8 microwatts. However, the life of pacemakers varies from 6 to 12 years. Therefore, replacing a new battery requires costly and painful surgery. Moreover, the leads of pacemakers and defibrillators have a considerable risk of failure [83], [84]. Hence, efficient use of heartbeat energy can be a suitable solution for such problems [85]–[89].

- Blood pressure: Blood pressure is a measure of the force of blood against the walls of arteries as the heart pumps it around the body. Energy harvesting from blood pressure via the piezoelectric effect to power embedded micro-sensors in the human body is a new technique to generate electricity. One challenge in capturing and quantifying electrical data in the human nervous system is the absence of durable implantation interfaces that can sustain neural activity over extended periods. Recently, scientists have implanted microsensors deep within the brain to monitor local electrical and physiological data, which is then transmitted to an external interrogator. The energy collected through this process also holds potential for long-term applications [90]–[92].

- Breathing: Breathing is another source of human body power that can be used for energy recovery purposes. Breathing energy harvesting involves capturing and converting the energy generated during breathing into usable electrical energy. The concept is part of the broader field of energy harvesting, where ambient energy from various sources is harnessed for powering electronic devices or sensors. During inhalation and exhalation, air moves into and out of the respiratory system because of the pressure disparity between the lungs and the surrounding air. The maximum pressure of a breath can exceed atmospheric pressure by 2 %, which equates to 1 W of power. The advantage of this source as a source of energy is its constant presence even during sleep, which can be a stable source of energy inversely, increasing the effort to inhale or exhale can impair the user's comfort, which is one of the disadvantages of this method. In addition, most breathing-based generators need breath masks that encumber the user. However, such masks are already in place for applications that can be used optimally, such as military pilots, astronauts, or hazardous materials handlers. Some individuals, like soldiers, miners, or those who regularly use anti-pollution masks, might find it useful if the mask could also function as a power source for portable or wearable electronic devices. In the context of WBANs, breathing energy harvesting aims to utilize the mechanical energy generated by chest or diaphragmatic movements during the breathing cycle. This mechanical energy can be converted into electrical energy using various technologies, such as piezoelectric materials, triboelectric nanogenerators, or other mechanisms [93], [94].

- Muscle movement: To harvest energy from a muscle movement, such as tapping a finger, muscle vibration near the throat, or stretching on a face, one must develop a different approach than the other methods mentioned above. This concept is part of the broader field of energy harvesting, where ambient or biomechanical energy is harnessed for various applications, including powering electronic devices or sensors. Muscle movement energy harvesting utilizes the mechanical strain or motion produced during muscle activities. This energy can be converted into electrical power using different technologies, such as piezoelectric materials, electromagnetic induction, or biomechanical transducers. In recent years, researchers proposed various methods to harvest this mechanical energy the novel methods consist of an alternating current generator that connects a piezoelectric wire at both ends of metal electrodes based on the annular tension and release. It is a promising method to harvest ideal energy from muscle movement, which can supply power for small and low-consumption medical devices [95], [96]. A summary of the latest representative research in human body energy harvesting in WBANs is shown in Table 2.

## IV. FUTURE RESEARCH DIRECTION

Researchers tried to find the optimal solutions for energy utilization in WBANs by enhancing existing methods or incorporating new technologies. Energy harvesting, as an increasing battery efficiency solution for powering WBAN nodes, is still in its infancy. To solve the current challenges in energy harvesting, effective solutions must be proposed. Moreover, considerable research must also be focused on future applications. This section highlights the benefits and challenges for future research trends in energy utilization in WBANs.

### A. Reinforcement Learning

RL in the context of energy harvesting in WBANs involves using RL algorithms to optimize the energy harvesting process and the overall performance of the WBAN. RL is a type of machine learning where an agent learns to make decisions by interacting with its environment and receiving feedback in the form of rewards. By applying RL to energy harvesting in WBANs, the system can learn to make intelligent decisions that improve energy efficiency, longer device lifetimes, and enhanced overall performance. This approach is especially beneficial in dynamic and unpredictable environments, where traditional rule-based strategies may fall short [97]–[101].

RL algorithms can dynamically allocate energy harvesting resources based on the varying energy needs of wearable devices in the WBAN. The system can learn to distribute energy efficiently among different devices to maximize overall performance. Moreover, RL enables the WBAN to adapt its energy harvesting strategies based on changing environmental conditions. For instance, the system can learn when to switch between different energy sources (e.g., solar, thermal, kinetic) depending on factors like ambient light, temperature, or user activity .

RL algorithms can model and adapt to user behavior patterns. The system can optimize energy harvesting schedules to align with increased movement or environmental energy availability by considering when and how users typically engage in physical activities. RL can be employed to manage the charging and discharging of batteries in wearable devices. The system learns optimal battery usage policies, considering factors such as device priority, energy storage capacity, and the criticality of different devices in the WBAN.In addition to device-level energy harvesting, RL can optimize communication strategies within the WBAN. Considering the devices' available energy and communication requirements, the system can learn when to activate communication modules.

RL enhances the WBAN's ability to adapt to device failures or unexpected changes in energy availability. The system learns to reconfigure itself in real-time to maintain essential functions and prolong the overall network lifetime. Multi-agent RL can be employed in scenarios where multiple WBANs interact (e.g., in healthcare settings with multiple patients). This allows the WBANs to learn collaborative energy-sharing and coordination strategies, optimizing the overall network performance. RL facilitates real-time decision-making for energy harvesting and management. The system can quickly adapt to sudden environmental or user behavior changes, ensuring timely and effective energy-related decisions. A summary of the future research direction of the RL in WBANs application is shown in Fig. 5.

TABLE II
SUMMARY OF SOME LATEST REPRESENTATIVE RESEARCH IN HUMAN BODY ENERGY HARVESTING IN WBAN

| Ref. | Source \ Technology | Method | Strategy | Main Work | Output |
|---|---|---|---|---|---|
| [54] | Human body | Analytical | Review | The text provides an overview of the dispersion and attributes of three main energy sources present in the human body, namely thermal energy, chemical energy, and mechanical energy | Several common demonstrations and uses of different energy harvesting technologies on the human body are listed accordingly |
| [63] | Glucose | Implementation | Reduce power consumption | Describes the development and deployment of a non-invasive glucose sensor node for WBAN | The patient is able to continuously monitor their glucose levels using a sensor, eliminating the need for frequent battery replacements due to the self-energy harvesting capability |
| [65] | Enzymatic Lactate | Experimental | Fabrication | A new type of enzymatic biofuel cell that does not require a membrane, and utilizes carbon nanotubes coated on carbon cloth as bioelectrodes | Significant stability was demonstrated by retaining 73% of the original catalytic activity even after a storage period of 15 days |
| [76] | Human body | Biomechanical energy harvesting | Review | Examine energy harvesting methodologies designed to capture biomechanical energy generated by human movement, including but not limited to foot impact, joint articulation, and upper limb motion | Methods for harvesting energy from the human body and their respective characteristics |
| [85] | Heartbeat | Electrostatic energy harvesters | Modeling and simulation | Exploring a three-dimensional characteristic through the application of a nonlinear state-space methodology in the context of optimal energy harvesting from heartbeats | The power generation has significantly increased to 35.038 w at the same scale when compared to the previous rate |
| [86] | Heartbeat | Quasi-Concertina structure design | Modeling and simulation | A new and innovative mechanism has been developed for capturing the energy generated by heartbeats to supply power to leadless implantable pacemakers | By employing a Teflon electret charged to 1000 V, the apparatus is capable of producing an average power output of 10.06 W when subjected to a vibration amplitude of 0.25 m/s2 at its resonant frequency |
| [90] | Blood pressure | Piezoelectric effect | Simulation | Harvesting energy from the arterial blood pressure using the piezoelectric effect to power micro-sensors implanted in the human brain | The theoretical framework is confirmed through the utilization of a Multiphysics 3D-FEA simulation to assess the generated power across various load resistances |
| [92] | Human Arm Movement | PMN-35PT | Simulation | Conducting a comprehensive examination of three Relaxer (1-x)PMN-xPT ceramic compositions, with a focus on the impact of composition on piezoelectric, dielectric, and electromechanical properties | The suggested approach exhibited consistent precision in measuring systolic blood pressure |

## B. Hybrid solution

Hybrid solution refers to the integration of multiple energy harvesting techniques or sources to enhance overall energy efficiency and reliability. This approach aims to leverage the strengths of different energy harvesting methods to address the limitations of individual techniques. Renewable energy harvesting systems have been focusing on their potential use in self-powered smart WBAN systems. Using a hybrid energy harvesting method can obtain significant improvements in network lifetime. Using a single energy source has different restrictions on the functioning and reliability of a sensor node to energy harvesting, especially in the WBAN's devices connected to the human body, supplying energy with only one source is very risky and unreliable. Hybrid energy harvesting technology is proposed to solve the energy-insufficiency problem of a single energy harvester. Therefore, the researchers focused on multiple sources energy harvesting to exploit the full potential of energy harvesting in WBANs. This solution provides extra power and improves reliability for wireless nodes. Moreover, the hybrid solution includes multiple transduction mechanisms, mixed materials, and structures to improve energy harvesting efficiency. Consequently, hybrid energy harvesters can be classified into multi-source, and single-source harvesters with hybrid agents [102], [103].

## C. Deep Learning

Deep learning algorithms, a subset of machine learning, are other methods that can be used in WBAN. Deep learning is at a higher level of accuracy than ever before and helps consumer electronics meet user expectations.

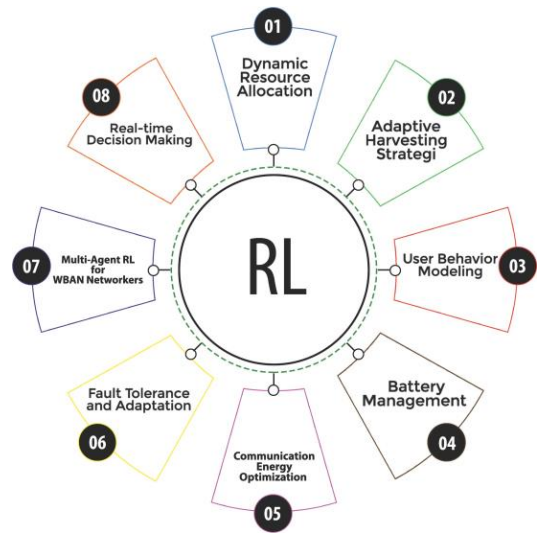

Fig. 5. A summary of the future research direction of the RL in WBANs application.

Therefore, this method is one of the researchers favorite topics. Deep Learning techniques can be effectively applied to energy harvesting in WBANs to optimize energy management, improve efficiency, and enhance overall system performance. Deep learning involves using advanced neural network models to optimize and enhance the efficiency of energy harvesting systems. Utilization of DL in energy harvesting in WBANs can be used of neural networks for predicting energy availability from various sources such as solar, thermal, and kinetic energy, or train deep learning models on historical data to forecast future energy levels based on environmental conditions and user activities. Dynamic resource allocation, optimized charging and discharging strategies, harvesting energy from various sources, user behavior modeling, anomaly detection and fault tolerance, communication energy optimization, and multi-modal data fusion are other ways deep learning can be employed in the context of energy harvesting in WBANs [104], [105].

### D. Development wearable/implantable devices

Due to society's increasing need for health devices, communities' demand for the use of WBAN is felt more than before. Therefore, the development of wearable and implantable devices in the body, especially in reducing dimensions and sizes, can consume less energy. Many researchers are researching and building in this area, but considering energy issues, it still requires more work. The development of wearable and implantable devices in the context of Energy Harvesting in WBANs involves creating devices that can capture and utilize energy from the body or its surroundings to power electronic components. In the applications of wearable device, integrating energy-harvesting technologies into fabrics for wearable devices such as Smart Clothing and using motion-based energy harvesting for devices that monitor physical activity as activity trackers can be two main goals to improve the devices. In implantable devices, biomechanical energy harvesting implants that place devices within the body that harness energy from movements or vibrations and biochemical energy harvesting by tapping into the body's biochemical processes for energy can be a challenging topic of this research direction. Main Challenges in this topic include miniaturization, biocompatibility, and efficiency. In this regard, developing compact devices suitable for implantation, ensuring materials used in implants are safe for the body, and improving energy harvesting efficiency to meet electronic components' power needs can be important research directions [106]–[108].

### E. Distributed Optimization

Distributed optimization involves decentralized approaches to optimize energy management and utilization across multiple devices or nodes within the network. This method includes decentralized energy harvesting control, collaborative energy sharing, game theory for energy allocation, distributed learning algorithms, consensus-based approaches, dynamic resource allocation, adaptive communication protocols, self-organizing networks, fault tolerance, resilience, Scalability, and Flexibility. Each of the mentioned methods can be a new topic in this field. Distributed optimization strategies in energy harvesting for WBANs are crucial for promoting autonomy, adaptability, and efficiency in managing energy resources. These approaches cater to the dynamic and decentralized nature of WBANs, making them well-suited for healthcare applications and other scenarios where energy efficiency is paramount [109]–[111].

## V. CONCLUSION

WBANs can play a prominent role in implementing health policy as an emerging phenomenon. One of the critical challenges in this field is the power supply of these networks to be convenient for the user. Also, this convenience will increase the audience using these devices. In this survey, we reviewed energy harvesting techniques in WBAN and categorized them. In this context, we provided recent literature on energy harvesting in WBAN. Furthermore, a classification of energy harvesting in WBAN is provided with appropriate references. Finally, we present future research direction. Our comprehensive can help researchers to find the features, similarities, differences, and implementation. Moreover, we believe that properly implementing WBAN in the medical and non-medical fields will improve the quality of life. In addition, energy energy harvesting methods mentioned in this paper can be used in other applications for small-scale energy.

12[74] Y. Zou, V. Raveendran, and J. Chen, "Wearable triboelectric nanogenerators for biomechanical energy harvesting," Nano Energy, vol. 77, p. 105303, 2020.

[75] J. Xiong et al., "Skin-touch-actuated textile-based triboelectric nanogenerator with black phosphorus for durable biomechanical energy harvesting," Nature communications, vol. 9, no. 1, p. 4280, 2018.

[76] Choi, Young-Man, Moon Gu Lee, and Yongho Jeon. "Wearable biomechanical energy harvesting technologies." Energies 10, no. 10 (2017): 1483.

[77] Zou, Yongjiu, Vidhur Raveendran, and Jun Chen. "Wearable triboelectric nanogenerators for biomechanical energy harvesting." Nano Energy 77 (2020): 105303.

[78] Liu, Mingyi, Feng Qian, Jia Mi, and Lei Zuo. "Biomechanical energy harvesting for wearable and mobile devices: State-of-the-art and future directions." Applied Energy 321 (2022): 119379.

[79] S. U. Ahmed et al., "Energy Harvesting through Floor Tiles," in 2019 International Conference on Innovative Computing (ICIC), 2019: IEEE, pp. 1-6.

[80] A. Proto, M. Penhaker, D. Bibbo, D. Vala, S. Conforto, and M. Schmid, "Measurements of generated energy/electrical quantities from locomotion activities using piezoelectric wearable sensors for body motion energy harvesting," Sensors, vol. 16, no. 4, p. 524, 2016.

[81] Houng, Hesmondjeet Oon Chee, Siti Sarah, S. Parasuraman, MKA Ahamed Khan, and I. Elamvazuthi. "Energy harvesting from human locomotion: gait analysis, design and state of art." Procedia Computer Science 42 (2014): 327-335.

[82] Ji, Sang Hyun, Yong-Soo Cho, and Ji Sun Yun. "Wearable core-shell piezoelectric nanofiber yarns for body movement energy harvesting." Nanomaterials 9, no. 4 (2019): 555.

[83] Y. Zhang, B. Lu, C. Lü, and X. Feng, "Theory of energy harvesting from heartbeat including the effects of pleural cavity and respiration," Proceedings of the Royal Society A: Mathematical, Physical and Engineering Sciences, vol. 473, no. 2207, p. 20170615, 2017.

[84] N. Li et al., "Direct powering a real cardiac pacemaker by natural energy of a heartbeat," Acs Nano, vol. 13, no. 3, pp. 2822-2830, 2019.

[85] Pourahmadi-Nakhli, Meisam, Bahareh Sasanpour, Mostafa Mahdavi, and Mohsen Sharifpur. "Optimal Heartbeat Energy Harvesting using Electrostatic Energy Harvesters." Energy Technology (2023): 2300569.

[86] Maamer, Bilel, Nesrine Jaziri, Mohamed Hadj Said, and Fares Tounsi. "High-displacement electret-based energy harvesting system for powering leadless pacemakers from heartbeats." Archives of Electrical Engineering 72, no. 1 (2023).

[87] Hu, Christopher, Kamran Behdinan, and Rasool Moradi-Dastjerdi. "PVDF energy harvester for prolonging the battery life of cardiac pacemakers." In Actuators, vol. 11, no. 7, p. 187. MDPI, 2022.

[88] Mariello, Massimo. "Heart Energy Harvesting and Cardiac Bioelectronics: Technologies and Perspectives." Nanoenergy Advances 2, no. 4 (2022): 344-385.

[89] Dong, Lin, Congran Jin, Andrew B. Closson, Ian Trase, Haley C. Richards, Zi Chen, and John XJ Zhang. "Cardiac energy harvesting and sensing based on piezoelectric and triboelectric designs." Nano Energy 76 (2020): 105076.

[90] A. Nanda and M. A. Karami, "Energy harvesting from arterial blood pressure for powering embedded micro sensors in human brain," Journal of Applied Physics, vol. 121, no. 12, 2017.

[91] L. Dong et al., "Multifunctional pacemaker lead for cardiac energy harvesting and pressure sensing," Advanced healthcare materials, vol. 9, no. 11, p. 2000053, 2020.

[92] Lifi, Houda, Amine Alaoui-Belghiti, Mohamed Lifi, Salam Khrissi, Naima Nossir, Yassine Tabbai, and Mohammed Benjell un. "Mechanical energy harvesting system from the human arm movement for continuous blood pressure measurement." International Journal of Sensors Wireless Communications and Control 12, no. 5 (2022): 352-368.

[93] Sahu, Manisha, Sugato Hajra, Sagar Jadhav, Basanta Kumar Panigrahi, Deepak Dubal, and Hoe Joon Kim. "Bio-waste composites for cost-effective self-powered breathing patterns monitoring: An insight into energy harvesting and storage properties." Sustainable Materials and Technologies 32 (2022): e00396.

[94] Li, Hui, Yannan Sun, Yujun Su, Ruihuan Li, Hongwei Jiang, Yingxi Xie, Xinrui Ding, Xiaoyu Wu, and Yong Tang. "Multiscale metal mesh based triboelectric nanogenerator for mechanical energy harvesting and respiratory monitoring." Nano Energy 89 (2021): 106423.

[95] Khalid, Salman, Izaz Raouf, Asif Khan, Nayeon Kim, and Heung Soo Kim. "A review of human-powered energy harvesting for smart electronics: Recent progress and challenges." International Journal of Precision Engineering and Manufacturing-Green Technology 6 (2019): 821-851.

[96] Panda, Swati, Sugato Hajra, Krystian Mistewicz, Pichaya In-na, Manisha Sahu, P. Mary Rajaitha, and Hoe Joon Kim. "Piezoelectric energy harvesting systems for biomedical applications." Nano Energy 100 (2022): 107514.

[97] Roy, Moumita, Dipanjana Biswas, Nauman Aslam, and Chandreyee Chowdhury. "Reinforcement learning based effective communication strategies for energy harvested WBAN." Ad Hoc Networks 132 (2022): 102880.

[98] Xu, Yi-Han, Jing-Wei Xie, Yang-Gang Zhang, Min Hua, and Wen Zhou. "Reinforcement learning (RL)-based energy efficient resource allocation for energy harvesting-powered wireless body area network." Sensors 20, no. 1 (2019): 44.

[99] Mohammadi, Razieh, and Zahra Shirmohammadi. "DRDC: Deep reinforcement learning based duty cycle for energy harvesting body sensor node." Energy Reports 9 (2023): 1707-1719.

[100] Roy, Moumita, Dipanjana Biswas, Nauman Aslam, and Chandreyee Chowdhury. "Reinforcement learning based effective communication strategies for energy harvested WBAN." Ad Hoc Networks 132 (2022): 102880.

[101] Gupta, Arti, and Vijay Kumar Chaurasiya. "Reinforcement learning based energy management in wireless body area network: A survey." In 2019 IEEE Conference on Information and Communication Technology, pp. 1-6. IEEE, 2019.

[102] H. Liu, H. Fu, L. Sun, C. Lee, and E. M. Yeatman, "Hybrid energy harvesting technology: From materials, structural design, system integration to applications," Renewable and sustainable energy reviews, vol. 137, p. 110473, 2021.

[103] Olatinwo, Damilola D., Adnan M. Abu-Mahfouz, and Gerhard P. Hancke. "A hybrid multi-class MAC protocol for IoT-enabled WBAN systems." IEEE Sensors Journal 21, no. 5 (2020): 6761-6774.

[104] Lv, Mingsong, and Enyu Xu. "Deep learning on energy harvesting iot devices: Survey and future challenges." IEEE Access 10 (2022): 124999-125014.

[105] Islam, Sahidul, Jieren Deng, Shanglin Zhou, Chen Pan, Caiwen Ding, and Mimi Xie. "Enabling fast deep learning on tiny energy-harvesting IoT devices." In 2022 Design, Automation & Test in Europe Conference & Exhibition (DATE), pp. 921-926. IEEE, 2022.

[106] Maity, Shovan, Debayan Das, and Shreyas Sen. "Wearable health monitoring using capacitive voltage-mode human body communication." In 2017 39th Annual International Conference of the IEEE Engineering in Medicine and Biology Society (EMBC), pp. 1-4. IEEE, 2017.

[107] Dao, Nhu-Ngoc. "Internet of wearable things: Advancements and benefits from 6G technologies." Future Generation Computer Systems 138 (2023): 172-184.

[108] Kiourti, Asimina, and Konstantina S. Nikita. "A review of in-body biotelemetry devices: Implantables, ingestibles, and injectables." IEEE Transactions on Biomedical Engineering 64, no. 7 (2017): 1422-1430.

[109] Arafat, Muhammad Yeasir, Sungbum Pan, and Eunsang Bak. "Distributed energy-efficient clustering and routing for wearable IoT enabled wireless body area networks." IEEE Access 11 (2023): 5047-5061.

[110] Zhumayeva, Merey, Kassen Dautov, Mohammad Hashmi, and Galymzhan Nauryzbayev. "Wireless energy and information transfer in WBAN: A comprehensive state-of-the-art review." Alexandria Engineering Journal 85 (2023): 261-285.

[111] Xin, Ge, Fengye Hu, Zhuang Ling, Shun Na, and Chi Jin. "Dynamic Scheduling for Minimizing Age Penalty in Resource-Constrained Classified WBANs with Energy Harvesting." IEEE Sensors Journal (2023).